# Quantum-inspired terahertz spectroscopy with visible photons

Mirco Kutas[1,2,*], Björn Haase[1,2], Jens Klier[1], Daniel Molter[1] and Georg von Freymann[1,2]

[1]*Fraunhofer-Institute for Industrial Mathematics ITWM, Fraunhofer-Platz 1, 67663 Kaiserslautern, GERMANY*
[2]*Department of Physics and Research Center OPTIMAS, Technische Universität Kaiserslautern (TUK), 67663 Kaiserslautern, GERMANY*
*Correspondence: Mirco Kutas ([mirco.kutas@itwm.fraunhofer.de](mirco.kutas@itwm.fraunhofer.de))*

**Abstract**

Terahertz spectroscopy allows for identifying different isomers of materials, for drug discrimination as well as for detecting hazardous substances. As many dielectric materials used for packaging are transparent in the terahertz spectral range, substances might even be identified if packaged. Despite these useful applications, terahertz spectroscopy suffers from the still technically demanding detection of terahertz radiation. Thus, either coherent time-domain-spectroscopy schemes employing ultrafast pulsed lasers or continuous-wave detection with photomixers requiring two laser systems are used to circumvent the challenge to detect such low-energetic radiation without using cooled detectors. Here, we report on the first demonstration of terahertz spectroscopy, in which the sample interacts with terahertz idler photons, while only correlated visible signal photons are detected – a concept inspired by quantum optics. To generate these correlated signal-idler photon pairs, a periodically poled lithium niobate crystal and a 660 nm continuous-wave pump source are used. After propagating through a single-crystal nonlinear interferometer, the pump photons are separated from the signal radiation by highly efficient and narrowband volume Bragg gratings. An uncooled scientific CMOS camera detects the frequency-angular spectra of the remaining visible signal and reveals terahertz-spectral information in the Stokes as well as the anti-Stokes part of collinear forward generation. Neither cooled detectors nor expensive pulsed lasers for coherent detection are required. We demonstrate spectroscopy on the well-known absorption features in the terahertz spectral range of α-lactose monohydrate and para-aminobenzoic acid by detecting only visible photons.

**Introduction**

Spectroscopy in the terahertz frequency range has shown its applicability to numerous tasks in the last decades[1-4]. The most prominent measurement principle in this specific part of the electromagnetic spectrum is time-domain spectroscopy (TDS) with ultrashort pulses from femtosecond lasers[5]. As these lasers are the main cost drivers for terahertz systems, more cost-efficient sources are currently investigated[6-9]. However, laser sources and optics in the visible range are rather cheap and highly developed. In order to use these also for experiments in other spectral regions, quantum sensing and imaging have recently become popular schemes[10-12]. One of the most promising applications seems to be quantum spectroscopy based on the effect of induced coherence without induced emission[13,14]. Since its first demonstration[15], quantum spectroscopy has gained attention especially in the infrared spectral range[16,17]. Furthermore, quantum spectroscopy in the terahertz frequency range has previously been demonstrated, using a single-crystal interferometer to measure the absorption of a periodically poled lithium niobate (PPLN) crystal[18] and a two-crystal setup measuring the linear properties of 4% magnesium oxide doped lithium niobate (LiNbO$_3$) crystal[19]. However, in these two experiments only the crystal material itself could be investigated, since the terahertz radiation was not coupled out of the nonlinear crystal, which is a crucial requirement for measurements of external samples. Thus, we demonstrate terahertz spectroscopy of α-lactose monohydrate and para-aminobenzoic acid by only detecting visible photons as a proof-of-concept experiment.

**Experimental setup**

The schematic of the experimental setup shown in Fig. 1 is an advanced version of our previously presented nonlinear interferometer[20]. As nonlinear media 1-mm-long PPLN with a poling period of either 220 µm or 200 µm are used. The crystals are illuminated by a 660 nm continuous-wave pump laser that is coupled into the interferometer via a volume Bragg grating and generates correlated pairs of visible signal and terahertz idler photons. After the crystal, an off-axis parabolic mirror (OAP) with a through-hole is placed, separating the terahertz photons from the pump and signal photons. Due to the high refractive index of lithium niobate in the terahertz frequency range[21], the scattering angles of the terahertz radiation are large and even terahertz photons corresponding to collinearly emitted signal photons can have an angle with respect to its associated signal radiation[20]. This allows to spectrally separate these wavelengths purely depending on the angle while only a small part of the terahertz radiation is lost through the hole in the mirror. Using an OAP with a through-hole instead a beam splitter for separation has the additional advantage that no signal and pump losses occur. After the second pass through the nonlinear crystal, the pump photons are efficiently filtered from the signal photons by volume Bragg gratings acting as highly efficient notch filters (filter section in Fig. 1). The remaining signal radiation is focused through a transmission grating to observe a frequency-angular spectrum on the sCMOS camera. To receive an interference of the signal radiation, the path-length difference is changed by moving the reflective mirror ($M_i$) with a piezoelectric linear stage.

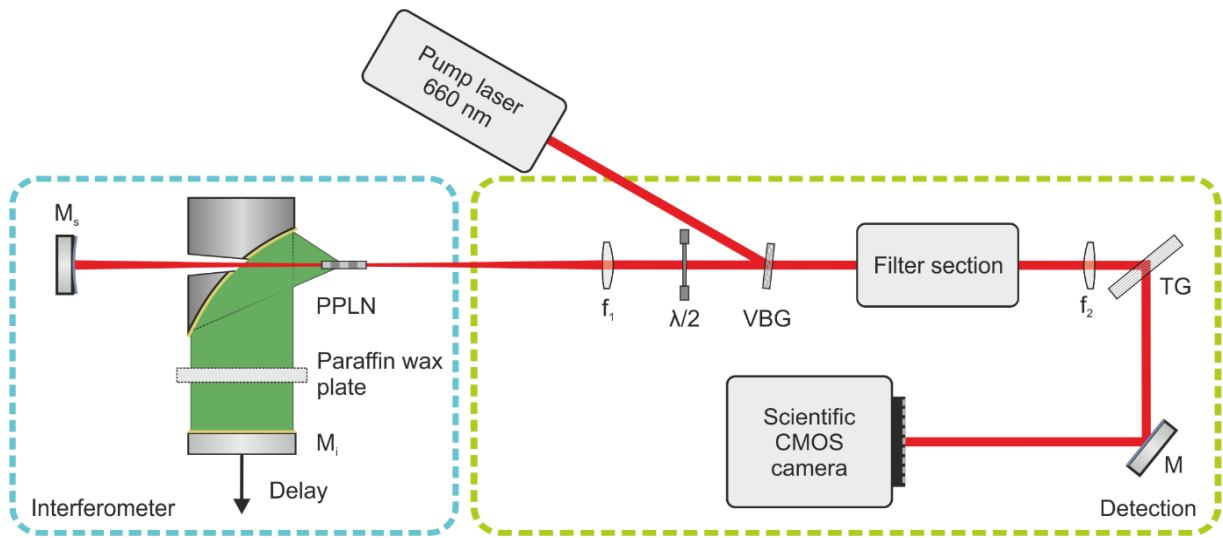

**Fig. 1 Experimental setup.** The 1-mm-long PPLN crystals are pumped by a continuous-wave laser with a wavelength of 660 nm generating correlated pairs of signal and terahertz photons. After the crystal the terahertz radiation is separated by an OAP with a through-hole and afterwards reflected at a moveable mirror $M_i$. Pump and generated signal photons are reflected at $M_s$ directly back into the crystal. After the second pass the pump radiation is filtered from the signal radiation by three volume Bragg gratings (VBG). To obtain a frequency-angular spectrum on the sCMOS camera, the signal radiation is focused through a transmission grating (TG).

**Frequency-angular spectra**

The acquired frequency-angular spectra show the characteristic tails for terahertz radiation in the Stokes and anti-Stokes region (see Fig. 2a). These signal photons are either generated by spontaneous parametric down-conversion (SPDC) or conversion of thermal terahertz photons, as reported earlier[22,23]. As a consequence, the Stokes- and the anti-Stokes part only differ by higher count rate and higher visibility of the Stokes part, due to the contribution of SPDC. For that reason, we limit the evaluation in this work to the Stokes part (see Fig. 2b). To obtain interference, frequency-angular spectra are recorded for several path-length differences and a fast Fourier transform (FFT) of the waveform measured of each individual pixel is carried out. In Fig. 2c the maximum amplitude of the individual FFT of each pixel in the range between 0.2 THz and 1.0 THz is shown. The highest amplitudes can be observed for the collinear forward regions at about 0.52 THz. Due to the large scattering angle of the terahertz radiation out of lithium niobate, an interference can only be observed in the collinear forward regions corresponding to the highest amplitudes[20]. However, in this representation the tails can still be observed, since pump-induced fluctuations in the signal radiation are recorded as well.

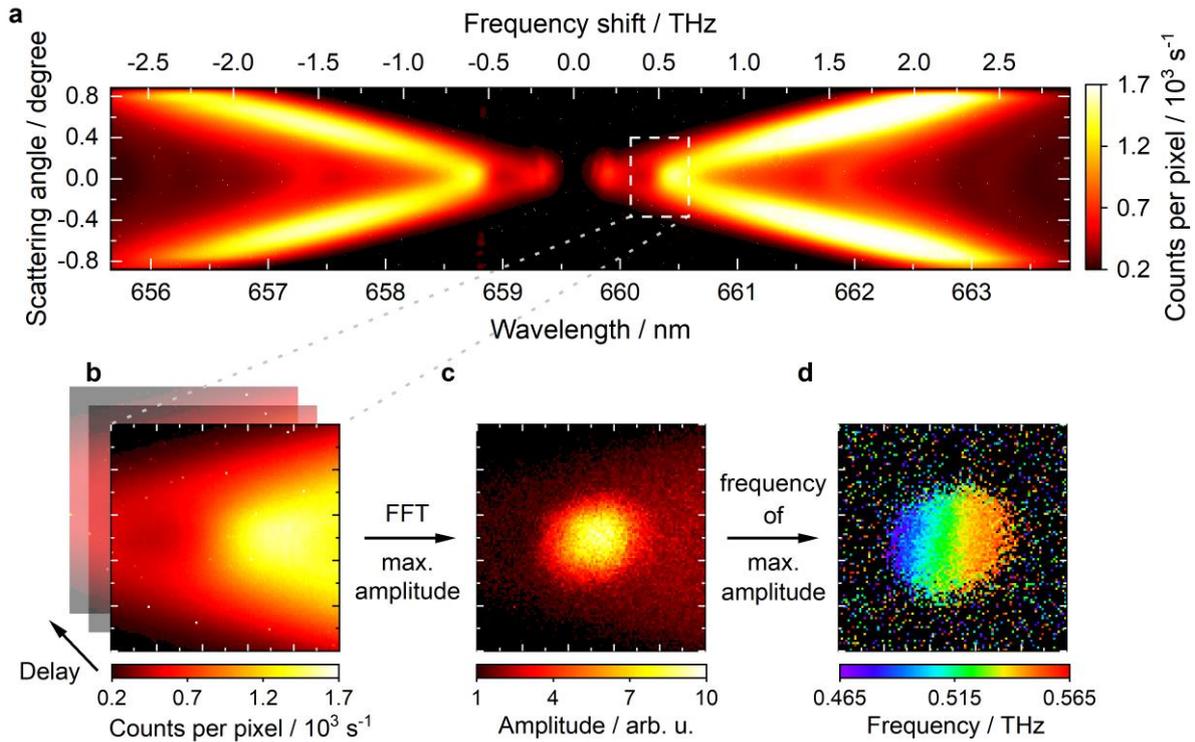

**Fig. 2 Frequency-angular spectrum. a** Frequency-angular spectrum for the used crystal with a poling period of $\Lambda$ = 220 μm. The image represents an averaging of 60 images with an exposure time of 500 ms each at a mirror position of $M_i$ = 0.2 mm. **b** Enlargement of the dashed area (100x100 pixels) in **a**. **c** Maximum amplitude of the FFT for each individual pixel and **d** corresponding frequency in the frequency range between 0.2 THz and 1.0 THz of the measured spectra (see **b**). The highest amplitudes are achieved in the collinear forward region. For a better display, the frequency range in **d** is limited to 0.465 THz to 0.565 THz.

Additionally, the frequency corresponding to the maximum amplitude in Fig. 2c is evaluated for each single pixel (see Fig. 2d). For a precise determination of the maximum of the FFT, zero filling was used. Since only the collinear areas of the forward-facing terahertz radiation show a frequency dependence, the frequency range is limited to 0.465 THz to 0.565 THz. The observed frequency of the interference increases with increasing frequency shift, leading to an accessible bandwidth of more than 100 GHz for a single crystal. While in the given experiment we only had access to crystal lengths of one millimeter, the bandwidth could be increased by shortening the crystal length. However, Fig. 2d shows, that pixel-wise spectral information is obtained from a measurement with a single crystal without varying external parameters and can be used to gain spectral information of samples in the terahertz frequency range.

**Terahertz spectroscopy**

To demonstrate the ability of quantum-inspired terahertz spectroscopy (QIS) by detecting only visible photons, well characterized samples are measured. As samples, paraffin wax plates are manufactured in which $\alpha$-lactose monohydrate or para-aminobenzoic acid are dissolved (see Methods). These substances are widely used for the demonstration of terahertz spectroscopy, as they show well-known characteristic absorption features. For terahertz spectroscopy, these additives are normally mixed with polytetrafluoroethylene powder and pressed under high pressure into tablets. Due to the rather large beam diameter of 2 inches in the given experiment, it was necessary to choose a different type of sample production, since such large molds were not available. Paraffin wax has proven to be an ideal carrier material, as it shows almost no absorption in the terahertz frequency range and has also a low refractive index[24] of about 1.54. The amount of the additives to the paraffin wax plate ranges between 0.0 g and 2.1 g. Considering the area of the plate of 107x80 mm$^2$ and assuming a homogeneous distribution, this corresponds to concentrations between 0 mg/mm$^2$ and 0.25 mg/mm$^2$, respectively.

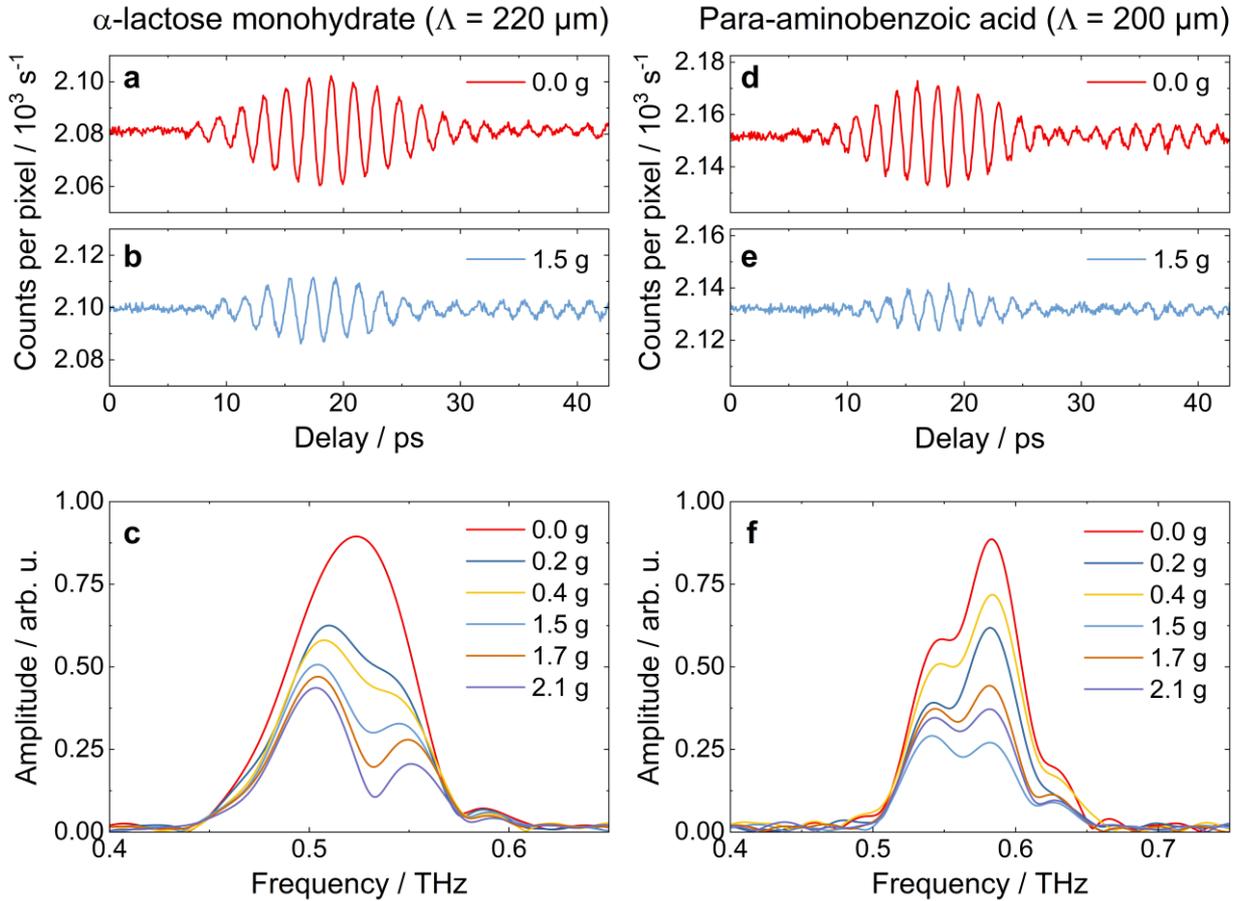

**Fig. 3 Waveforms and spectra.** Waveform of the used crystal with a poling period of **a,b** $\Lambda$ = 220 µm and **d,e** $\Lambda$ = 200 µm and a wax plate with an amount of **a,c** 0.0 g and **b,d** 1.5 g of the additive inserted in the terahertz path. The visbility of the measured waveforms without additive are about 1 % and after the interference envelope the response behavior of a water vapor absorption line at 0.56 THz is visible. Corresponding spectra of the used crystals with a poling period of **c** $\Lambda$ = 220 µm and **f** $\Lambda$ = 200 µm with wax plates of different amounts of **c** $\alpha$-lactose monohydrate and **f** para-aminobenzoic acid inserted in the terahertz beam path. For clarity, not all of the measured spectra are shown.

As one prominent absorption feature of $\alpha$-lactose monohydrate is at 0.53 THz and of para-aminobenzoic acid at 0.6 THz we use two different poling periods of 220 µm for $\alpha$-lactose monohydrate and 200 µm for para-aminobenzoic acid[8], respectively. In Fig. 3a and 3d the measured waveforms in the collinear forward region of the signal for the two crystals are shown where plates with only paraffin wax are already inserted into the terahertz path. For both crystals the maximum relative visibility is about 1% and some further interference after the main envelope is observable. This is mainly determined by the response behavior of water vapor in air having a rather weak absorption line at 0.56 THz. Additionally, in Fig. 3b and 3e the measured waveforms of paraffin wax plates with an amount of 1.5 g of the additives are shown. In comparison, these waveforms show a decreased amplitude and a different response behavior due to the ingredient. In Fig. 3c and 3f the frequency spectra of the interference in the collinear forward region for both poling periods are shown for different amounts of additive in the inserted paraffin wax plate. For the poling period of 220 µm, the paraffin wax plate without ingredient shows a peak at about 0.52 THz, in accordance to the calculated phase-matching conditions. The obtained spectra with α-lactose monohydrate added to the wax have a dip at 0.53 THz and the depth of this dip increases with increasing amount of the additive, as expected. In Fig. 3f the frequency spectrum of the crystal with a poling period of 200 µm is shown having a peak at about 0.58 THz, also in accordance to the phase-matching conditions. As the absorption of para-aminobenzoic acid is near to the observed spectral width of the interference, only a decrease of amplitude over almost the entire spectrum by an increased amount of additive is observed. In comparison to the other crystal, the spectrum does not show a clear envelope, as already for the reference plate several dips can be observed. These are caused by the water vapor absorption line at 0.56 THz. For the spectra in Fig. 3c, this absorption line is close to the noise level of the spectrum, while for Fig. 3f it is near to the maximum amplitude. As a result, the response behavior of the water line in the measured waveform is larger, but cannot be fully resolved as the available optical delay of 42.7 ps is rather small (corresponding to a maximum mirror displacement of 6.4 mm) and therefore directly limits the spectral resolution. This was also confirmed by simulations of the waveform considering water vapor absorption (not shown).

For validation of the presented method, extinction measurements of the samples have been performed with a standard TDS system: An area of 100x67 mm$^2$ of the paraffin wax plates is raster imaged. Thus, the homogeneity of the additive distribution in the sample can be evaluated. Figure 4c and 4d show two exemplary imaging results of paraffin wax plates with an amount of 1.5 g α-lactose monohydrate and para-aminobenzoic acid, respectively. For the sake of simplicity, only the peak-to-peak amplitude value of the time-domain waveform of each pixel is evaluated for imaging purposes. As can be seen, the distribution of the ingredient is not completely homogeneous.

In order to be able to compare the TDS measurement with the QIS measurement, only the area of the sample that is penetrated by the terahertz radiation in the QIS experiment (indicated by the dashed circle in Fig. 4c and 4d) is considered for the evaluation of the extinction for both experiments. Additionally, the emission characteristics of the terahertz radiation and the angular density of the terahertz radiation that penetrates a given area are taken into account (see Methods for details). Considering both, the extinction at the evaluated absorption line was calculated. As can be seen in Fig. 4a and 4b the measured extinctions by QIS show a good agreement with the extinctions measured with the standard TDS system reaching a coefficient of determination of $R^2 = 0.91$ for α-lactose monohydrate and $R^2 = 0.90$ for para-amino benzoic acid, respectively.

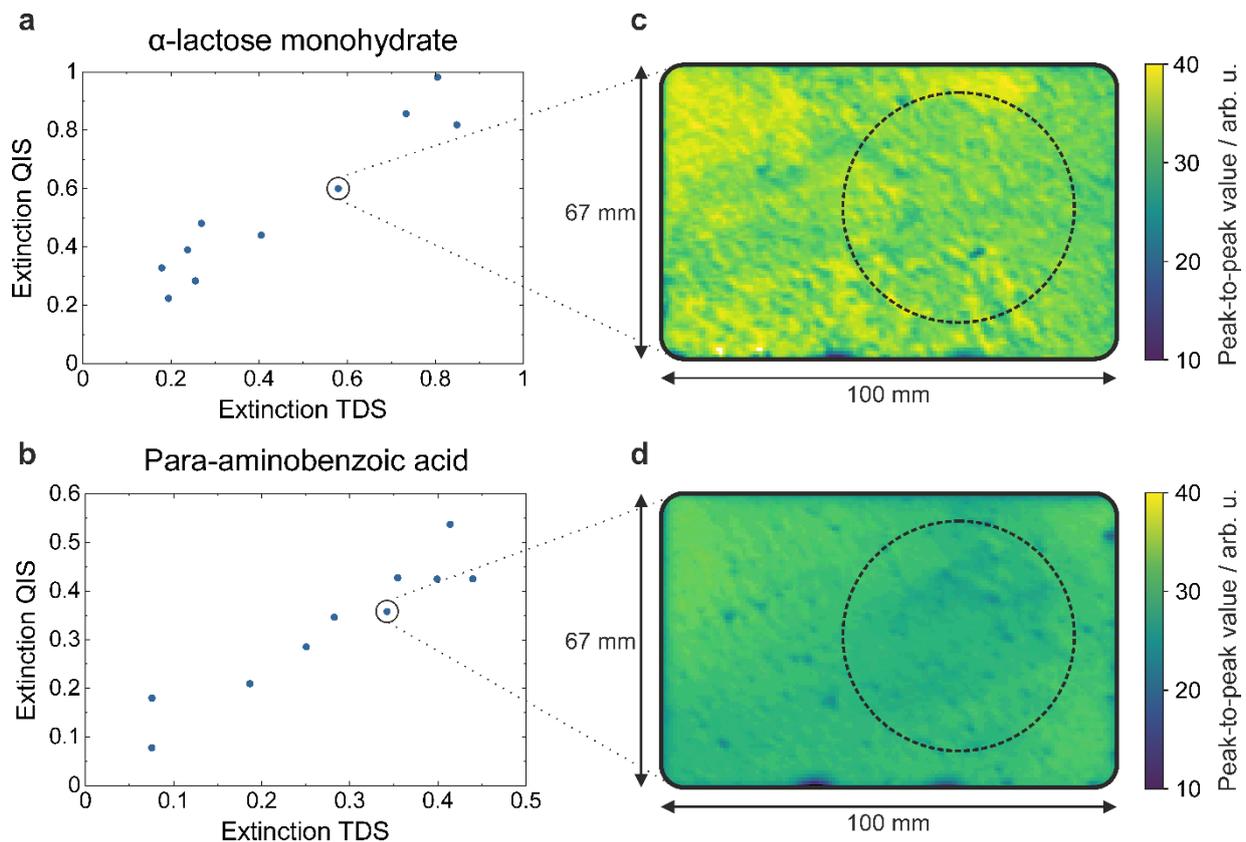

**Fig. 4 Comparison between QIS and conventional TDS.** Comparison of the extinction measured at the absorption frequency in the experiment and by a standard TDS system for wax plates with different amounts of **a** α-lactose monohydrate and **b** para-aminobenzoic acid. Raster image of a wax plate with an additive amount of 1.5 g **c** α-lactose monohydrate and **d** para-aminobenzoic acid measured by a conventional TDS system. The dashed circles indicate the area that is illuminated by the terahertz radiation in the QIS experiment and therefore evaluated for extraction of extinction values.

**Discussion**

With a determination coefficient of the extinction of at least 0.9, a good agreement between the experimental results achieved with the nonlinear interferometer and the conventional TDS system is observed. The frequency of the observed absorption lines in the QIS experiment agree well with literature values[8]. One of the reasons for the remaining deviation between the two measurements is expected to be due to the limited displacement of the used translation stage. As a result, information from the sample or water vapor absorption that is delayed compared to the envelope of the origin waveform is lost. This causes a low resolution of the measured feature, which cannot be improved by simply zero filling before the FFT is performed. Additionally, the region that is penetrated by the terahertz radiation may differ from the evaluated area of the raster image and the assumption of the angular density of the terahertz radiation is not precise enough. So far, the used LiNbO$_3$ crystals show a usable terahertz bandwidth of about 150 GHz and an interference visibility of more than 1 %. The limited bandwidth will be widened in the future, e.g., by employing shorter crystals while effects on increasing the visibility are currently investigated. Despite all this, the experiment shown is a proof-of-principle for spectroscopy in the terahertz frequency range without using terahertz detectors by only measuring visible photons. It shows the potential of determining sample properties in the terahertz frequency range without using expensive femtosecond lasers by replacing them with rather cheap and highly developed lasers in the visible range and silicon-based cameras.

**Conclusion**

In conclusion, we have demonstrated spectroscopy in the terahertz frequency range by only detecting visible photons. For the first time spectral information of an external sample in the terahertz frequency range is obtained using the concept of nonlinear interferometry. As a first proof-of-concept, the concentration of $\alpha$-lactose monohydrate and para-aminobenzoic acid solved in paraffin wax plates were determined. The estimated losses match the values measured by a standard TDS system fairly well. Although the achieved resolution and bandwidth are not yet comparable to classical terahertz measurement techniques, the presented demonstration of this concept is a first milestone towards multi-pixel terahertz spectral imaging.

## Methods

### Nonlinear interferometer

We use a single-crystal Michelson-like setup (see Fig. 1) having the advantage that just one crystal is used. The pump source is a linearly polarized, frequency-doubled, single-longitudinal-mode, continuous-wave laser (Cobolt Flamenco) at a center wavelength of 659.58 nm, providing a narrow bandwidth of less than 1 MHz at an average output power of up to 500 mW. The laser is coupled into the interferometer by a volume Bragg grating (BNF.660 from OptiGrate) acting as a mirror. In order to align the polarization of the pump radiation, a zero-order half-wave plate at the design wavelength of 670 nm is placed in the beam path, ensuring extraordinary pump polarization and type 0 phase matching in the PPLN crystal. The pump light is focused into the crystal by a lens with a focal length of 200 mm placed in the distance of its focal length. As nonlinear media for the generation of correlated photon pairs $LiNbO_3$ crystals with a dimension of 5×1×1 $mm^3$ and a quasi-phase matching structure with a poling period of 200 µm or 220 µm are used. Due to high absorption in the terahertz frequency range, longer PPLN crystals show no advantage. For the creation of signal and idler photons, energy and momentum conservation has to be fulfilled. The corresponding phase-matching considerations of $LiNbO_3$ are shown in detail in a previous work[23]. After the crystal, the terahertz photons are separated from the pump and signal photons by an OAP with a through-hole placed in the distance of its focal length of 50.8 mm behind the crystal. Because of the high refractive index of $LiNbO_3$ in the terahertz region, it is emitted under large scattering angles which allows a purely angle-dependent separation. To collect a large part of the terahertz photons, a 2-inch parabolic is used. After collimation, the terahertz radiation is reflected by a 2-inch gold-coated plane mirror that is placed on a coarse manual linear stage (resolution ± 5 µm) and an automated fine translation stage (Q-522.030 from PI; resolution ± 8 nm) with a maximum displacement of up to 6.4 mm and a minimum increment of 8 nm. Pump and signal photons are focused back into the nonlinear crystal by a concave mirror with a focal length of 100 mm placed at a distance of 200 mm to the crystal.

### Detection

The detection part in Fig. 1 of the experimental setup has remained unchanged from our previous work[20]. Pump and signal radiation are collimated by the former focusing lens and then enter the filter section. This essentially consists of three volume Bragg gratings (separated by a distance of 1 m from one another) acting as highly efficient notch filters. They efficiently reflect the pump photons while the down- and up-converted signal photons having a slightly shifted frequency are transmitted. To block divergent stray light originating from the volume Bragg gratings we use spatial filters. For the detection of the remaining signal radiation an uncooled scientific CMOS camera (Thorlabs Quantalux™ sCMOS camera) with a specified quantum efficiency of about 55% (at 660 nm) is used having a pixel size of 5.04×5.04 µm² and providing 2.1 megapixels with up to 87 dB dynamic range. At 20 °C, the specified pixel dark count rate and readout noise is about 20 counts per second and 1 $e^-$, respectively. The measured background illumination of the presented results is about 160 counts per second and pixel (combining dark count rate, remaining stray and ambient light). To be able to observe spectral information directly on the camera the signal photons are focused by a lens with a focal length of 400 mm through a highly efficient transmission grating (PCG-1908-675-972 from Ibsen photonics with 1908 lines per millimeter, efficiency greater than 94% at the used wavelength). This combination leads to a frequency-angular spectrum on the camera (see Fig. 2). All frequency-angular spectra are acquired with an illumination time of 500 ms and a pump power of 450 mW at room temperature. The first-order forward and backward generation in the Stokes as well as in the anti-Stokes region are observable. By the contribution of SPDC, the Stokes region receives a higher count rate than the anti-Stokes region, where only conversion of thermal photons occurs. Due to apertures of the wave plate and filters along the beam path, the spectrum shows a vertical limitation starting from a scattering angle at about 0.5 degrees. A more detailed description can be found in a previous work[23].

**Sample fabrication**

To fabricate the samples for spectroscopy, paraffin wax was melted at 125 °C. This temperature was chosen, because for wax smoothest surface structures are created at high temperatures when solidifying and it still remains below the flash point. In the melted wax, powdered para-aminobenzoic acid or $\alpha$-lactose monohydrate was added. In order to achieve a plate, the wax with the additive was poured onto an aluminum plate inside a 3D-printed frame with the inner dimensions of 107x80 mm$^2$ and a thickness of 2 mm. To get a smooth surface structure on both sides, another aluminum plate was pressed onto the frame because wax does not solidify evenly on the surface while cooling. Furthermore, that way the thickness of the wax plates is most homogeneous and the risk of air pockets at the surface is minimized. After twenty minutes of cooling by air, the plates are stable enough to be carefully detached from the aluminum surfaces and the frame.

**Measuring the paraffin wax plates**

To measure the amount of additive solved in the paraffin wax plate, a waveform is recorded for each plate. The recording of this waveform is analogues to our previous work[20]. For every measurement, the mirror $M_i$ is moved with a step size of 10 µm (change of the path length of 20 µm) over a distance of 6.4 mm leading to a maximum delay of 42.7 ps. At every mirror position, an image of the frequency-angular spectrum is recorded with an illumination time of 500 ms. To achieve a higher signal-to-noise ratio, this method is repeated 60 times. For every image the collinear forward region of the frequency-angular spectrum is integrated over an area of 25x20 pixels that is located around the maximum interference amplitude (see Fig. 2c). Averaging over all repetitions for every position of the mirror leads to the measured waveform (see Fig. 3a,b and 3d,e). Afterwards, zero-filling is applied and a FFT of the acquired waveform is performed. By integrating the acquired spectrum in the region of the absorption and reference, the extinction with respect to the wax-only plate is calculated.

**Measuring the paraffin wax plates with a standard TDS system**

The paraffin wax plates are measured with a conventional TDS system. Therefore, the sample plate is raster-imaged by the TDS system with a spot size of about 1 mm and a step size of 0.7 mm in order to cover the whole plate. In Fig. 4c and 4d, the peak-to-peak amplitude values of the time-domain waveforms, measured with the TDS system, of plates with an amount of 1.5 g of $\alpha$-lactose monohydrate and para-aminobenzoic acid are shown. As one can see, there are small variations of the sample homogeneity inside the plate. These inhomogeneities are mainly caused by the fact that α-lactose monohydrate and para-aminobenzoic acid are not fully soluble in paraffin wax and because of their higher densities, the additives are not distributed homogeneously in the volume. To verify the QIS measurements, the extinction at the evaluated absorption line was calculated in the region of interest that is penetrated by the terahertz radiation in the given experiment (indicated by the dashed line in Fig. 4c and 4d). To do so, the time-domain spectra in the region of interest are summed up pixel-wise using a weighting factor for each pixel to account for the emission of the terahertz radiation as well as the OAP dependence. To take the radial characteristics of the generated terahertz radiation into account, an exponential decay from the center is assumed. Due to the geometry of the OAP, the angular density of the terahertz radiation penetrating a given area is different. The part closer to the crystal is able to collect larger angles of the emitted terahertz radiation what was considered by a parabolic dependence. Afterwards, a FFT of the achieved spectrum is performed and evaluated by integrating the acquired spectrum in the region of the considered absorption. By relating this value to the wax-only measurement, the extinction in the region of interest for each paraffin wax plate is calculated.

**Acknowledgments:** We thank P. Bickert for helpful discussions during this project.

**Funding:** This project was funded by the Fraunhofer-Gesellschaft within the Fraunhofer Lighthouse Project Quantum Methods for Advanced Imaging Solutions (QUILT).

**Author contributions:** M.K., B.H. and D.M. designed the experiment. M.K. and B.H. manufactured the samples and carried out the experiment. J.K. carried out the measurements with the standard TDS system. G.v.F. and D.M. supervised this research. All authors discussed the results and contributed to the writing of the manuscript.

**Data Availability:** All experimental data and any related experimental background information not mentioned in the text are available from the authors upon reasonable request.

**Conflict of interest:** The authors declare that they have no conflict of interest.